# Regression test of various versions of STRmix

Jo-Anne Bright[1], Judi Morawitz[1], Duncan Taylor[2,3] and John Buckleton[4]

1. ESR, P.B.92021 Auckland, New Zealand.
2. Forensic Science SA, 21 Divett Place, Adelaide, SA 5000, Australia
3. School of Biological Sciences, Flinders University, GPO Box 2100 Adelaide SA, Australia 5001
4. University of Auckland, Department of Statistics, Auckland, New Zealand

**Abstract**

STRmix™ has been in operational use since 2012 for the interpretation of forensic DNA profiles.  During that time incremental improvements have been made to the modelling of variance for single and composite peaks, and drop-in.  The central element of the algorithm has remained as the Markov chain Monte Carlo (MCMC) based on the Metropolis-Hastings algorithm.

Regression experiments are reported comparing versions V2.3 to V2.9.

There is a high degree of similarity in *LR*s for true contributors across multiple versions of STRmix™ that span 7 years of development (from V2.3 to V2.9). This is due to the fact that the underlying core models of STRmix™ that are used to describe DNA profile behaviour have remained stable across that time.

**Introduction**

STRmix™ has been in operational use since 2012 for the interpretation of forensic DNA profiles.  During that time incremental improvements have been made to the modelling of variance for single and composite peaks, and drop-in.  The central element of the algorithm has remained as the Markov chain Monte Carlo (MCMC) based on the Metropolis-Hastings algorithm.  Some major changes to the biological and statistical models are listed below:

- In version 2.3 the allele and stutter variances were, themselves, allowed to vary.
- In version 2.3 the MCMC stepping was changed to the current Gaussian walk.
- In version 2.3 the HPD interval was expanded to include theta (which was also now able to be specified as a distribution) and MCMC variability.
- In version 2.3 the presence of relatives of the POI in the population was introduced into *LR* calculations.
- In version 2.3.07 a change was made to disallow the chains to pass through the plane of symmetry that exists between two donors of similar template.
- In version 2.4 forward stutter modelling was introduced.
- In versions 2.6, 2.7 and 2.8 drop-in modelling was changed and refined.
- In version 2.6 generalised stutter was introduced.
- In version 2.6 limits were placed on the level of variability that could be exhibited by peaks with very low expected heights (Taylor Quantum Effect).
- In versions 2.7 and 2.8 a limit was placed on the size of the HPD interval generated as a result of allele proportion uncertainty (the limit based on population size and then later the minimum resampled count).

- In version 2.7 the variance of a composite peak was changed from a weighted average of the contributing peaks to the shifted log normal model.
- In version 2.7 the starting template values per contributor within the MCMC were changed from a 1000 rfu to dynamic values based on the profile.
- In version 2.7 the locus specific amplification efficiency variance was added to the model as a parameter.
- In version 2.8 a restriction was placed on potential stutter peaks so that if drop-in was a more favourable explanation, then it would be invoked instead of forcing a stutter.
- In version 2.9 a restriction was added that meant an undersized stutter peak could not become more favourable by adding fluorescence of a type with higher peak height variability.

These changes are incremental rather than revolutionary and often designed to deal with infrequently encountered situations, to expand the types of profiles that could be analysed, or reduce the amount of data pre-processing that was required. We would expect a regression test of a modern version against a legacy version to give similar results for all but the fringe cases that the modelling improvements were designed to address. Again, this would mean that testing and quality scrutiny is, to a major extent, cumulative rather than confined to one version.

**Method**

Experiment 1. The original input files and parameters from the study "A sensitivity analysis to determine the robustness of STRmix™ with respect to laboratory calibration" [1] were used. These files had been set up for V2.5 and had retained forward and back stutter but not double back and 2 bp back stutter. Both V2.5 and V2.9 can, therefore, interpret these input files.

The sample set containing mixtures (N = 71) with varying numbers of contributors (35 × NoC = 2, 36 × NoC = 3) and varying degrees of degradation was selected from the PROVEDIt set [2]. These had been profiled using Applied Biosystems™ GlobalFiler™ using 29 cycles of PCR and analysed on a 3500 genetic analyser. These were analysed using STRmix™ V2.5.10 and V2.9.1 using the same settings. $LR$s were calculated with propositions that included the POI and the remaining contributors as unknown under $H_1$ and all contributors unknown under $H_2$. Mixtures were tested against the true donors ($H_1$ tests) and a set of known non-donors ($H_2$ tests).

Experiment 2. A total of 18 validation profiles from Forensic Science South Australia, that had been analysed in versions of STRmix™ from versions V2.3 to V2.9 were compiled and $LR$s compared. A description of these mixtures appears in Table 1.

$LR$s were calculated with propositions that included all known DNA donors in $H_1$ and all unknowns in $H_2$. These propositions maximise the potential effects of variability in modelling. The recorded values were the sub-source point estimate (no HPD interval) of the unified and stratified (across Caucasian, Asian, and Aboriginal sub-populations from [3]) $LR$.

A description of the mixtures used and some summary data for experiment 2 appears in table 1. Labelling the data in the graph led to a cluttered view where the vertical spread of $\log_{10}LR$ became difficult to perceive. We invite the reader to use the average $\log_{10}LR$ to identify the mixtures if desired. Samples V1 to V15, and V17 are GlobalFiler™, Sample V16 is Profiler Plus™.

Table 1. A description of the mixtures used in experiment 2, the average $\log_{10}LR$ and the number of versions (including some developmental versions) tested.

| Mixture label, NoC, ratios, and total template | Average $\log_{10}LR$ | Number of versions tested |
|---|---|---|
| V1 2p, 1:1 400pg | 42.1 | 11 |
| V2 2p, 2:1 400pg | 53.6 | 11 |
| V3 2p 1:1 50pg | 37.2 | 11 |
| V4 2p 2:1 50pg | 34.4 | 11 |
| V5 2p 10:1 50pg | 28.1 | 11 |
| V6 3p 1:1:1 50pg | 27.5 | 11 |
| V7 2p 1:1:1 50pg | 31.9 | 11 |
| V8 3p 3:2:1 50pg | 31.7 | 11 |
| V9 3p 3:2:1 50pg | 30.9 | 11 |
| V10 2p 20:1 100pg | 47.9 | 11 |
| V11 2p 20:1 100pg | 43.4 | 11 |
| V12 2p 50:1 50pg | 30.6 | 11 |
| V13 2p 100:1 100pg | 30.6 | 11 |
| V14 1p 100pg | 13.9 | 6 |
| V15[1] - GF 1p 100pg | 14.8 | 6 |
| V16[2] - PP 1p 100pg | 9.3 | 6 |
| V17[3] - GF + PP (multi-kit) 1p 100pg | 14.6 | 5 |
| GF[4] 1-2 1-2p (GT=2p) 100pg | -11.3 | 4 |

Experiment 3

A comparison of the $\log_{10}LR$s for 32 mixtures using STRmix™ versions 2.8 and 2.9 was undertaken. The mixtures used appear in Table 2.

Table 2. The mixtures used in experiment 3.

| 1000pg 10-5-2-1 | 400pg_100-1 - 2 PCR | 400pg_4-3-2-1 - GS |
|---|---|---|
| 100pg - 10-5-2-1 | 100pg_1-1-1 - 3p to 4p | 400pg_20-1 - GS |
| 200pg - 20-10-10-1 | 100pg_1-1-1 | 400pg_200-1 - GS |
| 400pg - 50-25-10-1 | 200pg_1-1-1 - GS | 400pg_20-10-5-1 - Assume C123 |
| 3500xl2 - 50pg - 1-1 | 200pg_10-5-2-1 - GS | 50pg_1-1 - 2p |
| 50pg_1-1-1 - 3p | 400pg_1-1 | 50pg_5-1 - 2 2p to 3p |
| 50pg_5 -1 - 2 PCR | 400pg_1-1-1 - GS | 50pg_1-1-1 - 3p to 4p |
| 200pg_50 -1 - 2 PCR | 400pg_3-2-1 - 3p to 4p | 400pg_1-1 - 2p to 3p |

---

[1] This is an artificially created single source sample made to contain the most common genotypes in Australia
[2] This is the same sample as V15
[3] This is the two samples V15 and V16 run together using the multi-kits function
[4] This is a sample treated using the varNoC function with options NoC = 1 or 2. Ground truth is NoC = 2. This is a 2p GF mixture with a major and minor. The person being compared is not a contributor and is excluded under 1p and favours exclusion under 2p

| 200pg_50-1 - 2 2p to 3p | 400pg_3-2-1 | 400pg_1-1-1 - 3p to 4p |
| --- | --- | --- |
| 400pg_20-10-5-1 - 2Assume C234 - 4p to 5p | 400pg_4-3-2-1 - Assume C123- 4p to 5p | 400pg_1-1-1 |
| 400pg_100-1 - 2 2p to 3p | 400pg_4-3-2-1 - Assume C123 | |

**Results**

Experiment 1. The sub-source $LR$s for the 178 $H_1$ true tests in the two versions are plotted in Figure 1. The sub-source $LR$s for the 17,572 $H_2$ true tests in the two versions are plotted in Figure 2.

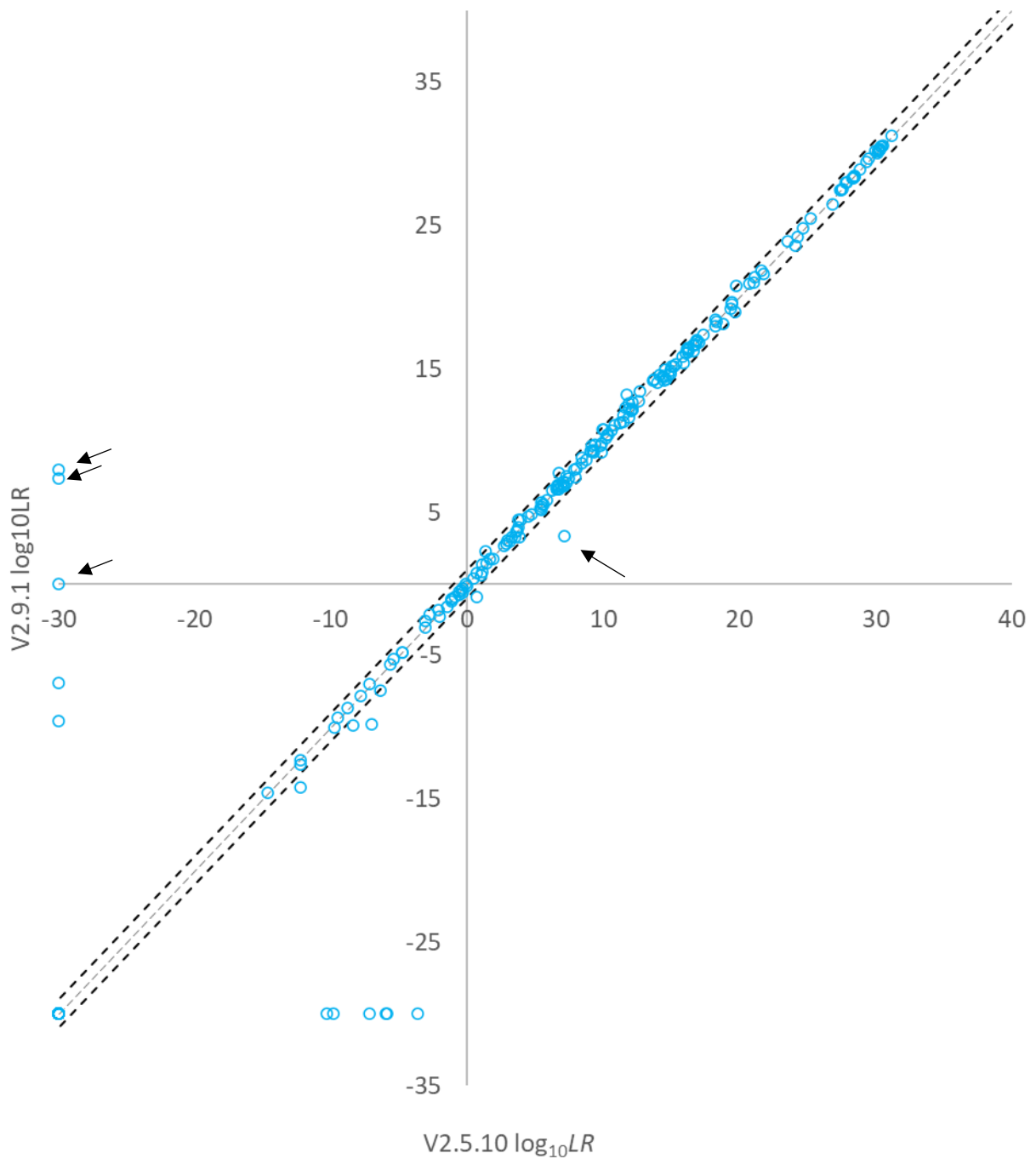

Figure 1. *x-y* scatter plot of the results for the 178 $H_1$ true tests for Experiment 1. Four data points are signified by an arrow. These are examined in more detail within Table S1 of the Appendix. The $x = y$ and the $x = y \pm 1$ are shown.

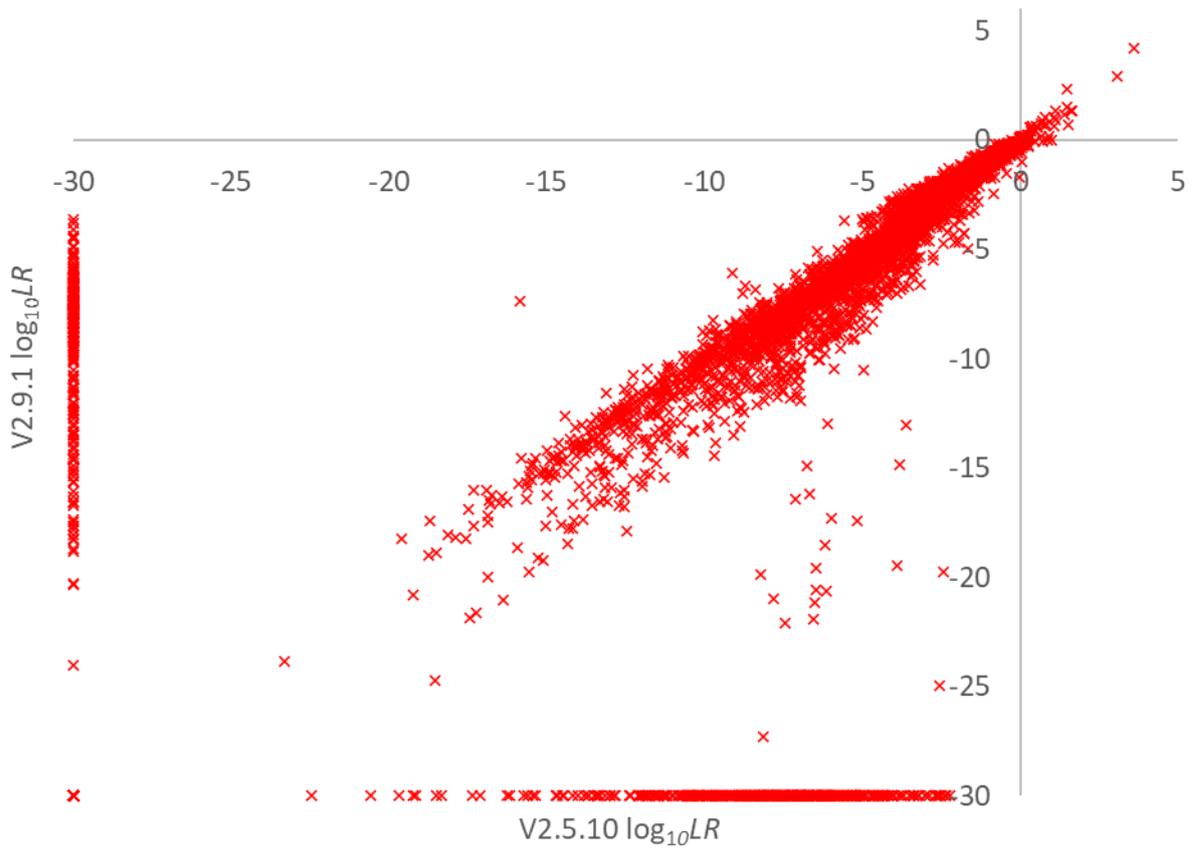

Figure 2. *x-y* scatter plot of the results for the 17,572 $H_2$ true tests for Experiment 1.

The four data points examined in detail all require a drop-in under $H_1$. Additional notes on these four data points appears in the Appendix, Table S1.

Experiment 2:

The results of the retrospective comparison of *LR*s for the 18 validation profiles are shown in Figure 3.

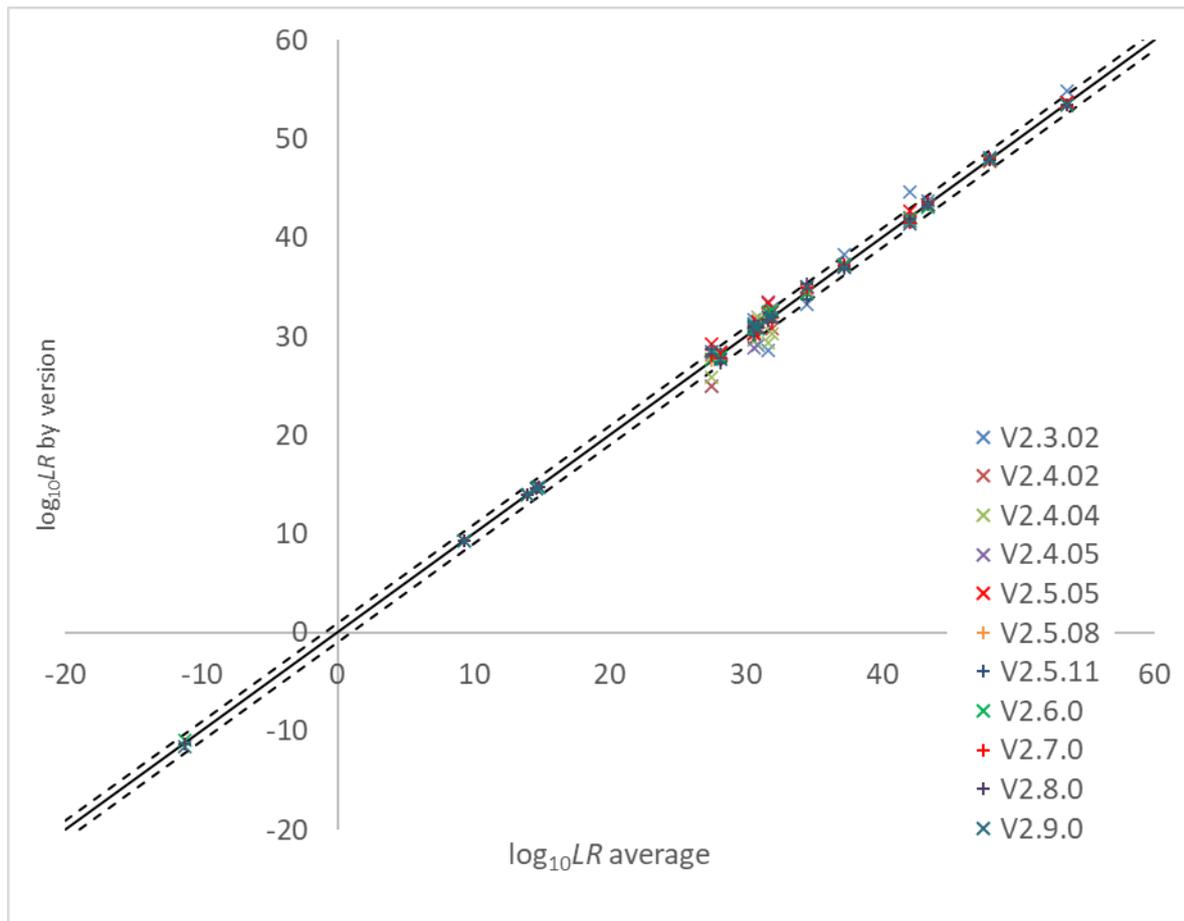

Figure 3: comparison of *LR*s generated for 18 validation profiles tested in STRmix™ versions from V2.3 to V2.9 for Experiment 2. The $x = y$ and the $x = y \pm 1$ are shown.

Experiment 3.

The results for Experiment 3 appear in Figure 4.

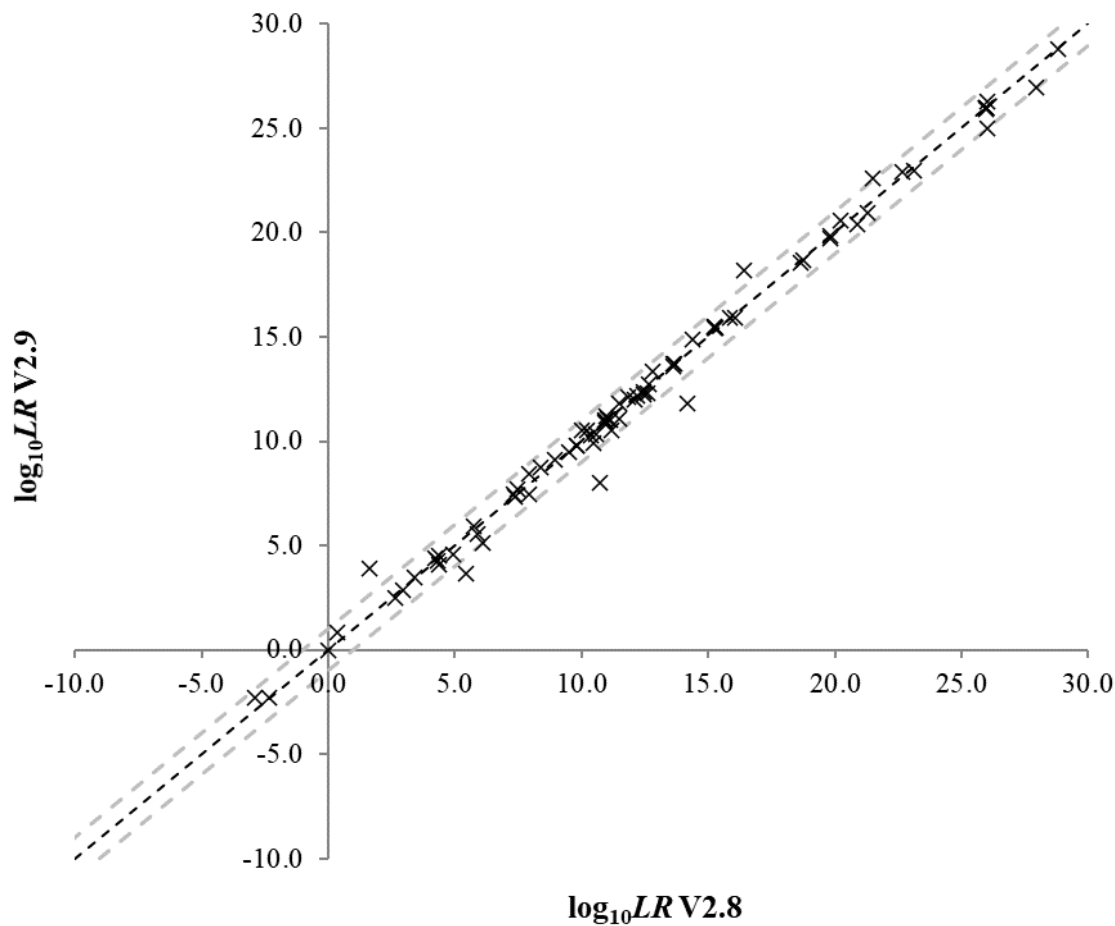

Figure 4. A plot of $\log_{10}LR$ for the mixtures described in Experiment 3 in STRmix™ V2.8 and V2.9. The $x = y$ and the $x = y \pm 1$ are shown.

**Conclusion**

There is a high degree of similarity in *LR*s for true contributors across multiple versions of STRmix™ that span 7 years of development (from V2.3 to V2.9). This is due to the fact that the underlying core models of STRmix™ that are used to describe DNA profile behaviour have remained stable across that time. Changes to modelling that have occurred in STRmix™ are generally to:

- Improve the performance of STRmix™ in modelling fringe cases,
- Expand the level of functionality in STRmix™ to accommodate a greater pool of DNA profiles of varying behaviour.

Taking this concept further, the main bulk of validation work that has been carried out on previous versions of STRmix™ are as applicable for current versions as they were for the version in which they were originally generated. The only exception to this is for those profiles that fall into the categories of fringe cases or previously uninterpretable profiles that the modelling improvements were targeted to address.

**Appendix**

Table S1. The four data points examined in detail.

| Evidence | True donor tested | Log$_{10}$LR V2.5.10 | Log$_{10}$LR V2.9.1 |
|---|---|---:|---:|
| L1_K42K43_47 | K42 | 7.1 | 3.4 |
| L1_K44K45_27 | K44 | -30[5] | 8.0 |
| L1_K42K43_4 | K42 | -30 | 7.4 |
| L1_K46K47K48_41 | K47 | -30 | 0.0 |

Sample 1:

Examination of the comparison of L1_K42K43_47 with the true donor K42

Table S2. The *LR*s by locus for the two comparisons using V2.5.10 and V2.9.1

| | V2.5.10 | | | V2.9.1 | | |
|---|---|---|---|---|---|---|
| Locus | Pr(E\|Hp) | Pr(E\|Hd) | LR | Pr(E\|Hp) | Pr(E\|Hd) | LR |
| D3S1358 | 1.61E-02 | 1.19E-02 | 1.35E+00 | 1.66E-02 | 1.23E-02 | 1.35E+00 |
| vWA | 6.08E-04 | 2.54E-04 | 2.40E+00 | 6.80E-04 | 2.62E-04 | 2.60E+00 |
| D16S539 | 1.32E-03 | 3.65E-04 | 3.63E+00 | 1.36E-03 | 3.86E-04 | 3.52E+00 |
| CSF1PO | 3.07E-02 | 2.63E-02 | 1.17E+00 | 3.05E-02 | 2.64E-02 | 1.16E+00 |
| TPOX | 5.11E-02 | 4.59E-03 | 1.11E+01 | 4.85E-02 | 4.31E-03 | 1.13E+01 |
| Yindel | | | | | | |
| D8S1179 | 1.67E-02 | 2.46E-03 | 6.78E+00 | 2.05E-02 | 2.43E-03 | 8.43E+00 |
| D21S11 | 1.09E-03 | 1.36E-03 | 7.96E-01 | 1.16E-03 | 1.39E-03 | 8.32E-01 |
| D18S51 | 1.61E-03 | 3.48E-04 | 4.61E+00 | 1.75E-03 | 3.55E-04 | 4.94E+00 |
| DYS391 | | | | | | |
| D2S441 | 5.65E-03 | 2.75E-03 | 2.05E+00 | 6.02E-03 | 2.80E-03 | 2.15E+00 |
| D19S433 | 1.68E-03 | 7.63E-05 | 2.21E+01 | 1.56E-03 | 7.41E-05 | 2.11E+01 |
| TH01 | 6.17E-03 | 7.61E-03 | 8.11E-01 | 6.25E-03 | 7.58E-03 | 8.25E-01 |
| FGA | 8.62E-06 | 5.81E-06 | 1.48E+00 | 8.35E-06 | 5.46E-06 | 1.53E+00 |
| D22S1045 | 1.20E-04 | 4.13E-05 | 2.90E+00 | 1.16E-04 | 4.07E-05 | 2.85E+00 |
| D5S818 | 4.20E-02 | 4.91E-02 | 8.56E-01 | 4.52E-02 | 4.81E-02 | 9.38E-01 |
| D13S317 | 1.23E-02 | 2.79E-02 | 4.42E-01 | 1.13E-02 | 2.77E-02 | 4.08E-01 |
| D7S820 | 1.38E-02 | 3.75E-03 | 3.69E+00 | 1.33E-02 | 3.77E-03 | 3.52E+00 |
| SE33 | 4.06E-05 | 7.19E-06 | 5.65E+00 | 1.62E-09 | 2.15E-06 | 7.51E-04 |
| D10S1248 | 1.53E-03 | 2.17E-03 | 7.06E-01 | 1.45E-03 | 2.29E-03 | 6.30E-01 |
| D1S1656 | | | | | | |
| D12S391 | 8.24E-03 | 5.28E-03 | 1.56E+00 | 8.01E-03 | 5.40E-03 | 1.48E+00 |
| D2S1338 | 1.86E-03 | 3.50E-04 | 5.31E+00 | 1.81E-03 | 3.40E-04 | 5.31E+00 |
| Sub-Sub-Source | LR | 2.77E+07 | | LR | 4.58E+03 | |
| Sub-source | LR | 1.38E+07 | | LR | 2.29E+03 | |

---

[5] -30 is used when the *LR* is 0 and hence the log cannot be calculated.

The largest difference is at SE33 (highlighted in Table 2).

Table S3 The peak heights and alleles at the SE33 locus above AT.

|      | Allele | Height | mwt    |
|------|--------|--------|--------|
| SE33 | 16     | 179    | 354.24 |
| SE33 | 17     | 1274   | 358.27 |
| SE33 | 20     | 118    | 370.45 |
| SE33 | 21     | 1157   | 374.5  |
| SE33 | 22     | 111    | 378.65 |
| SE33 | 25.2   | 127    | 392.53 |

Table S4. The genotypes of the two true donors to L1_K42K43_47

| Reference ID | SE33 | |
|--------------|------|------|
| K42          | 25.2 | 25.2 |
| K43          | 17   | 21   |

The 22 peak within the evidence file is either a large forward stutter or a drop-in.

Table S5. The genotype probability distributions for the SE33 locus (note the drop-in peaks for V2.5 have been inferred from the genotype set and are not part of the genotype pdf file as they are in V2.9)

| SE33 V 2.5.10 | | | |
|---|---|---|---|
| Genotype C1 | Genotype C2 | Drop-in | Weight |
| [17,21] | [22,25.2] |  | 0.807682139 |
| [17,21] | [16,25.2] | 22 | 0.140840426 |
| [17,21] | [17,25.2] | 22 | 0.017502459 |
| [17,21] | [25.2,25.2] | 22 | 0.012024451 |
| [17,21] | [21,25.2] | 22 | 0.011995398 |
| [17,21] | [-1,25.2] | 22 | 0.005616737 |
| [17,21] | [20,25.2] | 22 | 0.002662012 |
| [17,21] | [16,22] | 25.2 | 0.001194819 |
| [17,21] | [16,17] | 22, 25.2 | 3.30E-04 |
| [17,21] | [22,22] | 25.2 | 8.13E-05 |
| SE33 V2.9.1 | | | |
| [17,21] | [22,25.2] |  | 0.998367575 |
| [17,21] | [16,22] | 25.2 | 4.97E-04 |
| [17,21] | [16,25.2] | 22 | 2.75E-04 |
| [17,21] | [17,22] | 25.2 | 2.70E-04 |
| [17,21] | [17,25.2] | 22 | 2.12E-04 |
| [17,21] | [21,25.2] | 22 | 1.94E-04 |
| [17,21] | [25.2,25.2] | 22 | 9.56E-05 |
| [17,21] | [-1,25.2] | 22 | 4.02E-05 |
| [17,21] | [21,22] | 25.2 | 3.45E-05 |
| [17,21] | [22,22] | 25.2 | 1.41E-05 |

The genotype combinations that align with the two known contributors are highlighted in Table S5. The difference in *LR* is that STRmix™ V2.9 is allowing drop-in for 22 more rarely.

Sample 2:

Examination of the comparison of L1_K42K43_4 with the true donor K42

The *LR* for TH01 is 0 for V2.5. The true donors are 8,9.3 and 9,9.3

Table S6 The peak heights and alleles at the TH01 locus above AT.

|      | Allele | Height | mwt    |
|------|--------|--------|--------|
| TH01 | 7      | 149    | 191.23 |
| TH01 | 8      | 1229   | 195.31 |
| TH01 | 9      | 147    | 199.3  |
| TH01 | 9.3    | 1681   | 202.39 |

TH01 stutter regression is $SR = 0.007541 + 0.001577 \times LUS$. Hence the expected SR for the 8 allele is about 0.020. The observed SR is 0.12. STRmix™ V2.5 has all the weight on genotype 7,9 for the minor. V2.9 allows the 7 to be drop-in.

Table S7. The genotype probability distributions for the TH01 locus

| V2.5.10     |             |         |          |
|-------------|-------------|---------|----------|
| Genotype C1 | Genotype C2 | Drop-in | Weight   |
| [8,9.3]     | [7,9]       |         | 1        |
| V2.9.1      |             |         |          |
| [8,9.3]     | [7,9]       |         | 0.999764 |
| [8,9.3]     | [9,9.3]     | 7       | 1.51E-04 |
| [8,9.3]     | [7,9.3]     | 9       | 5.12E-05 |
| [8,9.3]     | [7,8]       | 9       | 3.35E-05 |

The difference is that STRmix™ V2.9 is allowing drop-in for the 7 whereas V2.5 is not. Neither model the 7 as a large stutter.

Sample 3:

Examination of the comparison of L1_K44K45_27 with the true donor K44

The *LR* is 0 at D13S317 in V2.5.10

Table S8. The genotypes of the two true donors to L1_K44K45_27 at D13S317

| Reference ID | D13S317 |    |
|--------------|---------|----|
| K44          | 8       | 8  |
| K45          | 12      | 13 |

Table S9. The peak heights and alleles at the D13S317 locus above AT.

|  | Allele | Height | mwt |
|---|---|---|---|
| D13S317 | 8 | 849 | 210.83 |
| D13S317 | 12 | 259 | 227.02 |
| D13S317 | 13 | 410 | 231.1 |
| D13S317 | 14 | 120 | 235.19 |

Table S10. The genotype probability distributions for the D13S317 locus

| V2.5.10 | | | |
|---|---|---|---|
| Genotype C1 | Genotype C2 | Drop-in | Weight |
| [8,13] | [12,14] |  | 0.455 |
| [8,12] | [13,14] |  | 0.199 |
| [12,13] | [8,14] |  | 0.126 |
| [8,14] | [12,13] |  | 0.081 |
| [12,14] | [8,13] |  | 0.070 |
| [13,14] | [8,12] |  | 0.069 |
| V2.9.1 | | | |
| [8,13] | [12,14] |  | 0.465 |
| [8,12] | [13,14] |  | 0.181 |
| [12,13] | [8,14] |  | 0.136 |
| [8,14] | [12,13] |  | 0.083 |
| [13,14] | [8,12] |  | 0.071 |
| [12,14] | [8,13] |  | 0.060 |
| [8,13] | [8,12] | 14 | 8.80E-04 |
| [8,8] | [12,13] | 14 | 7.92E-04 |
| [12,13] | [8,8] | 14 | 5.14E-04 |
| [8,12] | [8,13] | 14 | 2.96E-04 |

The difference is that STRmix™ V2.9 is allowing drop-in whereas V2.5 is not.

Sample 4:

Examination of the comparison of L1_K46K47K48_41 with the true donor K47

The *LR* for FGA is 0 in V2.5.10.

Table S11. The genotypes of the three true donors to L1_K46K47K48_41 at FGA

| Reference ID | FGA | |
|---|---|---|
| K46 | 25 | 26 |
| K47 | 21 | 22 |
| K48 | 19 | 27 |

Table S12. The peak heights and alleles at the FGA locus above AT.

|  | Allele | Height | mwt |
|---|---|---|---|

| FGA | 19 | 112 | 247.57 |
|-----|----|-----|--------|
| FGA | 21 | 143 | 255.67 |
| FGA | 25 | 68  | 271.77 |
| FGA | 26 | 82  | 275.65 |

The analysis was run as NoC = 2 because that was the assignment from the epg. When run this way $H_1$ needs a drop-in of one of 19, 25, or 26

Table S13: The genotype probability distributions for the FGA locus (note the drop-in peaks for V2.5 have been inferred from the genotype set and are not part of the genotype pdf file as they are in V2.9)

| FGA V2.5 | | | |
|----------|----|----|----|
| Genotype C1 | Genotype C2 | Drop-in | Weight |
| [25,26] | [19,21] |    | 0.183 |
| [19,21] | [25,26] |    | 0.181 |
| [21,26] | [19,25] |    | 0.170 |
| [19,25] | [21,26] |    | 0.165 |
| [19,26] | [21,25] |    | 0.151 |
| [21,25] | [19,26] |    | 0.148 |
| [21,26] | [21,25] | 19 | 4.19E-04 |
| [-1,25] | [21,26] | 19 | 2.08E-04 |
| [25,25] | [21,26] | 19 | 9.03E-05 |
| [25,26] | [21,21] | 19 | 6.81E-05 |
| FGA V2.9 | | | |
| [19,21] | [25,26] |    | 0.199 |
| [21,26] | [19,25] |    | 0.181 |
| [25,26] | [19,21] |    | 0.162 |
| [19,25] | [21,26] |    | 0.160 |
| [19,26] | [21,25] |    | 0.155 |
| [21,25] | [19,26] |    | 0.145 |
| [21,26] | [25,26] | 19 | 9.21E-05 |
| [-1,21] | [19,26] | 25 | 1.60E-05 |
| [21,25] | [-1,19] | 26 | 8.23E-06 |
| [19,25] | [-1,26] | 21 | 5.15E-06 |

The difference is that STRmix™ V2.9 is allowing the 21,22 (it is aligned with the -1,21) by allowing drop-in of the 25. STRmix™ V2.5 has allowed drop-in of the 19 only.